\documentclass[preprint,showpacs,preprintnumbers,amsmath,amssymb]{revtex4}

\usepackage{graphicx}
\usepackage{dcolumn}
\usepackage{bm}

\begin{document}

\preprint{}

\title{Spatiotemporally-localized-stationary typical wave function \\satisfying Klein-Gordon equation with emergent mass}

\author{Agung Budiyono}
\affiliation{Institute for the Physical and Chemical Research, RIKEN, 2-1 Hirosawa, Wako-shi, Saitama 351-0198, Japan}

\date{\today}

\begin{abstract}

Starting from relativistic mass-less Madelung fluid, we shall develop a class of typical wave functions by imposing it to maximize Shannon entropy given its finite average quantum potential. We show that there is a class of solutions in which the wave function is spatiotemporally localized with finite spacetime support, uniformly moving hence stationary. It turns out that the quantum amplitude satisfies Klein-Gordon equation with emergent mass term proportional to the square root of average quantum potential. We show that there is physical time uncertainty which decreases as the mass increases. We also rederive the classical energy-momentum relation provided the de  Broglie-Einstein relation holds. In this case, the time uncertainty is proportional to the inverse of classical energy.  

\end{abstract}

\maketitle  

{\it Introduction: relativistic Madelung fluid} --- In Ref. \cite{AgungPRA1}, working with two dimensional Madelung fluid dynamics \cite{Madelung paper}, we asked the following question: {\it what is the typical wave functions of a single free particle given its energy}. By typical wave functions we meant the class of wave functions which maximizes Shannon entropy given some constraints \cite{Shannon entropy,Jaynes-MEP,Shore-Johnson-MEP}. We showed that there is a class of solutions in which the wave functions is self-trapped with finite-size spatial support, rotationally symmetric, spinning around its center, yet stationary. In this paper we shall carry out its relativistic extention for the non-rotating case. 

Starting from relativistic mass-less Madelung fluid dynamics, we shall develop a class of self-trapped spacetime wave function with finite-size spacetime support, uniformly moving thus stationary; and maximizes  Shannon entropy given its average quantum potential. We then show that its quantum amplitude satisfies Klein-Gordon equation with emergent mass term proportional to the square root of average quantum potential. We show that there is physical time uncertainty which decreases as the mass increases and eventually vanishes for infinite mass. We will also show that the ordinary classical energy-momentum relation holds if the de Broglie-Einstein relation is satisfied. In this case, the time uncertainty is shown to be proportional to the inverse of classical energy.    

First, the state of relativistic Madelung fluid in configuration spacetime $q^a=\{ct,x,y,z\}$, where $c$ denotes the velocity of light, is uniquely determined by a pair of fields $\{\rho(q),v^a(q)\}$. $\rho(q)$ is a scalar field kept to be normalized, thus the name quantum probability density. Whereas $v^a=dq^a/d\tau$, $\tau$ is proper time, is a four velocity vector field. From now on, the term wave function in this paper is used to refer to a pair of fields $\{\rho(q),v^a(q)\}$. Let us assume that their dynamics is governed by the following coupled covariant equations:
\begin{equation}
\frac{dv^a}{d\tau}=-\partial^aU,\hspace{3mm}
\frac{\partial\rho}{\partial\tau}+\partial_a\big(\rho v^a\big)=0.\hspace{2mm}
\label{covariant Madelung fluid dynamics}
\end{equation}
Here $U(q;\tau)$ is a scalar field called as quantum potential generated by quantum probability density as 
\begin{equation}
U(q;\tau)=\frac{\hbar^2}{2}\frac{\Box I}{I},
\label{covariant quantum potential}
\end{equation}
where $I(q)\equiv\rho^{1/2}(q)$ is quantum amplitude and $\Box\equiv (1/c^2)\partial_t^2-\partial_x^2-\partial_y^2-\partial_z^2$ is d'Alembertian operator. Due to its formal similarity with the Euler equation in hydrodynamics, the term on the right hand side of the left equation in Eq. (\ref{covariant Madelung fluid dynamics}), $F^a=-\partial^aU$, is called as quantum force field. 

\begin{figure}[tbp]
\begin{center}
\includegraphics*[width=6cm]{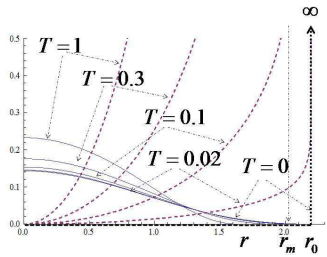}
\end{center}
\caption{The radial profile of the spherically symmetric self-trapped spatial quantum probability densities $\rho_s(r)$ (solid lines) and their corresponding quantum potentials $U_s(r)$ (dashed lines), for several small values of $T$. The quantum potential is shifted down so that its minimum is equal to zero. See text for detail. }
\label{rotationally symmetric 3D QPD-QP}
\end{figure}

{\it Typical -- self-traped quantum probability density with finite average quantum potential} --- Let us consider the following optimization problem: 
\begin{eqnarray}
\mbox{maximize:}\hspace{2mm}H[\rho]=-\int dq\hspace{1mm}\rho(q)\ln\rho(q),\nonumber\\
\mbox{constraint:}\hspace{2mm} \bar{U}=\int dq\hspace{1mm}U(q)\rho(q)\equiv U_{\mbox{\small tot}}.\hspace{0mm}
\label{MEP}
\end{eqnarray}
$H[\rho]$ is but Shannon entropy over the quantum probability density $\rho(q)$. This is the so-called maximum entropy principle, giving us the most probable, hence typical  $\rho(q)$ whose average quantum potential is $U_{\mbox{\small tot}}$. The solution of Eq. (\ref{MEP}) is given by \cite{Mackey-MEP}:
\begin{equation}
\rho(q)=\frac{1}{Z(T)}\exp\Big(-\frac{U(q)}{T}\Big),
\label{canonical QPD}
\end{equation}
where $T$ is the Lagrange constant below chosen to be non-negative and $Z(T)$ is a normalization factor. Notice that together with the definition of quantum potential given in Eq. (\ref{covariant quantum potential}), Eq. (\ref{canonical QPD}) comprises a differential equation for $U(q)$ or $\rho(q)$ subjected to the condition that $\rho(q)$ must be normalized. In term of $U(q)$, one has to solve the following nonlinear differential equation:
\begin{equation}
-\Box U=\frac{1}{2T}\partial^aU\partial_aU+\frac{4T}{\hbar^2}U. 
\label{covariant NPDE for U}
\end{equation}
One observes that the above differential equation is invariant under Lorentz transformation. Hence, given a solution $U(q)$ then a function $U(q')$, where $q'^a=\Lambda^a_{\hspace{1mm}b}q^b$ and $\Lambda^a_{\hspace{1mm}b}$ is Lorentz transformation, is also a solution of Eq. (\ref{covariant NPDE for U}). 

Let us develop a class of solutions in which the quantum probability density is being trapped by the quantum potential it itself generates \cite{AgungPRA1}. To do this, let us assume that there is an inertial frame so that the quantum probability density is separable into its spatial and temporal parts as follows:
\begin{equation}
\rho(q)=\rho_s({\bf x})\rho_t(t),\hspace{2mm}{\bf x}=\{x,y,z\}.
\label{separability}
\end{equation}
In this case, the quantum potential can be decomposed as $U(q)=U_s({\bf x})+U_t(t)$, where
\begin{equation}
U_s({\bf x})=-\frac{\hbar^2}{2}\frac{\partial_{\bf x}^2I_s}{I_s},\hspace{2mm}U_t(t)=\frac{\hbar^2}{2c^2}\frac{\partial_t^2I_t}{I_t}.
\label{decomposable quantum potential}
\end{equation}
Here, $I_i\equiv\rho_i^{1/2}$ with $i=s,t$; and $\partial_{\bf x}^2\equiv\partial_{\bf x}\cdot\partial_{\bf x}$ where $\partial_{\bf x}=\{\partial_x,\partial_y,\partial_z\}$. The condition of Eq. (\ref{separability}) is not Lorentz invariant so is the resulting class of solutions we are going to develop. Yet, we shall show that its nontrivial property will be shown to be Lorentz invariant. Equation (\ref{covariant NPDE for U}) can thus be collected as $\partial_{\bf x}^2U_s-(1/2T)\partial_{\bf x}U_s\cdot\partial_{\bf x}U_s-(4T/\hbar^2)U_s
=(1/c^2)\partial_t^2U_t-(1/2T)(\partial_tU_t)^2+(4T/\hbar^2)U_t=D,$
where $D$ is constant. Below we shall take the case when $D=0$. One thus has to solve the following decoupled pair of nonlinear differential equations:
\begin{eqnarray}
\partial_{\bf x}^2U_s-\frac{1}{2T}\partial_{\bf x}U_s\cdot\partial_{\bf x}U_s-\frac{4T}{\hbar^2}U_s=0,\nonumber\\
\frac{1}{c^2}\partial_t^2U_t-\frac{1}{2T}(\partial_tU_t)^2+\frac{4T}{\hbar^2}U_t=0.\hspace{2mm}
\label{decoupled NPDE}
\end{eqnarray}

\begin{figure}[tbp]
\begin{center}
\includegraphics*[width=6cm]{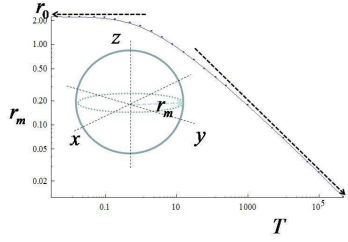}
\end{center}
\caption{The log-log plot of the radius $r_m$ of the three dimensional ball (inset) of spatial support againts $T$.}
\label{radius of quantum sphere vs T}
\end{figure}

To proceed, let us first discuss the spatial part by solving the upper differential equation in Eq. (\ref{decoupled NPDE}). To do this, let us confine ourself to a class of solutions which is spherically symmetric. In this case, one has to solve
\begin{equation}
\partial_r^2U_s(r)+\frac{2}{r}\partial_rU_s(r)-\frac{1}{2T}\big(\partial_rU_s(r)\big)^2-\frac{4T}{\hbar^2}U_s(r)=0, 
\label{rotationally symmetric NPDE for U}
\end{equation} 
where $r=\sqrt{x^2+y^2+z^2}$. Figure \ref{rotationally symmetric 3D QPD-QP} shows the numerical solutions of Eq. (\ref{rotationally symmetric NPDE for U}) with the boundary condition: $\partial_rU_s(0)=0$ and $U_s(0)=1$, for several small values of $T$. The reason for choosing small $T$ will be clear later. One can see that the spatial part of quantum probability density $\rho_s(r)$ is being trapped by its own self-generated quantum potential $U_s(r)$ \cite{AgungPRA1}. Moreover, there is a finite distance $r=r_m$ at which $U_s(r_m)$ is infinite so that $\rho_s(r_m)$ is vanishing. Hence, the self-trapped quantum probability density possesses only finite-size spatial support, $\mathcal{M}_s$, which takes the form of a three dimensional ball of radius $r_m$. See Fig. \ref{radius of quantum sphere vs T}. Fixing $\partial_rU_s(0)=0$, all these numerical observations are valid for any positive $U_s(0)$ \cite{extended version}.

Let us see what happens if one varies the parameter $T$. Figure \ref{radius of quantum sphere vs T} shows the values of $r_m$ as a function of $T$ obtained by numerically solving the differential equation of Eq. (\ref{rotationally symmetric NPDE for U}) with fixed boundary condition: $\partial_rU_s(0)=0$, $U_s(0)=1$. One can see that as we increase $T$, $r_m$ decreases, and eventually vanishes for infinite value of $T$. This shows that $\rho_s(r)$ is converging toward a delta function for infinite $T$. More interesting fact is seen for the opposite limit of vanishing $T$. One observes that $\lim_{T\rightarrow 0}r_m(T)=r_0$, where $r_0$ is finite. This fact suggests to us that at $T=0$, the spatial quantum probability density and thus its corresponding quantum potential are converging toward certain functions. Let us discuss this asymptotic situation in more detail. 

First, one can see in Fig. \ref{rotationally symmetric 3D QPD-QP} that as $T$ decreases, the quantum potential inside the spatial support is getting flatterer before becoming infinite at the boundary surface, $r=r_m(T)$. One may then guess that  at $T=0$, the quantum potential is perfectly flat inside the ball of spatial support and is infinite at its surface boundary, $r=r_0$. Guided by this guess, let us calculate the profile of the spatial quantum probability density for vanishing value of $T$, $\lim_{T\rightarrow 0}\rho_s(r;T)\equiv \rho_{s_0}(r)$. To do this, let us denote the assumed positive definite constant value of the quantum potential inside the support as $U_{s_0}$. Recalling the definition of spatial quantum potential given in Eq. (\ref{decomposable quantum potential}) and exploiting its spherical symmetry one has  
\begin{equation}
\partial_q^2 I_{s_0}(r)=\partial_r^2I_{s_0}(r)+\frac{2}{r}\partial_rI_{s_0}(r)=-\frac{2U_{s_0}}{\hbar^2}I_{s_0}(r),
\label{Schroedinger equation for a frozen ball}
\end{equation}
where $I_{s_0}\equiv\sqrt{\rho_{s_0}}$. The above differential equation must be subjected to the spatial boundary condition: $\rho_{s_0}(r_0)=I_{s_0}^2(r_0)=0$. Solving Eq. (\ref{Schroedinger equation for a frozen ball}) one has 
\begin{equation}
I_{s_0}(r)=A_{s_0}s_a(k_0r),
\label{quantum amplitude for a frozen ball}
\end{equation}
where $s_a$ is sinc function, $A_{s_0}$ is normalization constant and $k_{0}=\sqrt{2U_{s_0}/\hbar^2}$. The boundary condition implies $k_0r_0=\pi$. Figure \ref{rotationally symmetric 3D QPD-QP} shows that as $T$ decreases toward zero, $\rho_{s}(r;T)$ obtained by numerically solving Eq. (\ref{rotationally symmetric NPDE for U}) is indeed converging toward $\rho_{s_0}(r)$ obtained in Eq. (\ref{quantum amplitude for a frozen ball}). This observation thus confirms our guess that {\it at $T=0$, the spatial quantum potential is flat inside the spatial support $\mathcal{M}_s$ and is infinite at its surface boundary}. 

\begin{figure}[tbp]
\begin{center}
\includegraphics*[width=6cm]{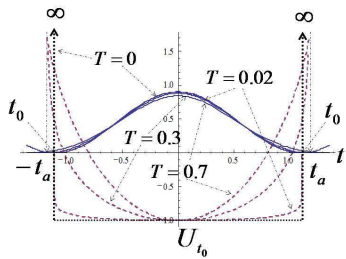}
\end{center}
\caption{The profile of self-trapped $\rho_t(t)$ (solid line) and its corresponding $U_t(t)$ (dashed line) for several small values of $T$. See text for detail.}
\label{tempo-creation-annihilation}
\end{figure}

Next let us discuss the temporal part of quantum probability density $\rho_t(t)$ by solving the lower differential equation in Eq. (\ref{decoupled NPDE}). In particular, we are interested to investigate the behavior of $\rho_t(t)$ at the limit $T\rightarrow 0$, if it exists. Figure \ref{tempo-creation-annihilation} shows the solution with the boundary: $\partial_tU_t(0)=0$ and $U_t(0)\equiv U_{t_0}=-1$. $\rho_t(t)$ and the corresponding $U_t(t)$ are plotted for several small values of $T$. One can again see similar phenomena with the spatial part, namely $\rho_t(t)$ is being self-trapped by the corresponding $U_t(t)$. The symmetry $\rho_t(-t)=\rho_t(t)$ is obvious and can be traced back to the differential equation of Eq. (\ref{decoupled NPDE}). One also sees that the support of $\rho_t(t)$ is finite given by the interval $\mathcal{M}_t=[-t_a,t_a]$. $t_a$ is smaller for smaller $T$ and is eventually converging toward a finite value for vanishing $T$, $\lim_{T\rightarrow 0}t_a\equiv t_0$. This again shows that at $T=0$, $\rho_t(t)$ will converge toward some function, $\lim_{T\rightarrow 0}\rho_t(t)\equiv\rho_{t_0}(t)$. Let us remark before proceeding that all the numerical observations concerning $U_t(t)$ in this paper is valid for any negative value of $U_t(0)$ \cite{extended version}. 

Proceeding in the same way as for the spatial part, let us calculate $\rho_{t_0}(t)$. To do this, first one observes that as $T$ is approaching zero, $U_t(t)$ is getting flatterer inside the support before becoming infinite at the boundary points, $t=\pm t_a(T)$. Again, let us guess that at $T=0$, the temporal quantum potential is perfectly flat inside the support $\mathcal{M}_t$, given by $U_{t_0}<0$; and is infinite at $t=\pm t_0$. Recalling the definition of $U_t(t)$ given in Eq. (\ref{decomposable quantum potential}), one has
\begin{equation}
\frac{\hbar^2}{2c^2}\partial_t^2I_{t_0}=U_{t_0}I_{t_0}.
\label{Schroedinger equation for time}
\end{equation}
Here $I_{t_0}\equiv \rho_{t_0}^{1/2}$. The above differential equation must be subjected to the boundary condition: $I_{t_0}(\pm t_0)=0$. Solving Eq. (\ref{Schroedinger equation for time}), one obtains:
\begin{eqnarray}
I_{t_0}(t)=A_{t_0}\cos(\omega_0 t). 
\label{temporal lowest state}
\end{eqnarray}
Here $A_{t_0}$ is a normalization constant and $\omega_0$ is related to the quantum potential as $\omega_0=\sqrt{(-2c^2U_{t_0}/\hbar^2)}$. The boundary imposes: $\omega_0t_0=\pi/2$. One finally sees in Fig. \ref{tempo-creation-annihilation} that  as one decreases $T$ toward zero, $\rho_t(t;T)$ obtained by solving the lower differential equation in Eq. (\ref{decoupled NPDE}) is indeed converging toward $\rho_{t_0}(t)$ given in Eq. (\ref{temporal lowest state}). {\it This again justifies our guess that at $T=0$, $U_{t}(t)$ is perfectly flat inside $\mathcal{M}_t$, and is infinite at $t=\pm t_0$}.   

{\it Stationary spacetime wave function and emergent mass} --- Hence, in total, at $T=0$, $\rho_0(r,t)\equiv\rho_{s_0}(r)\rho_{t_0}(t)$ satisfies the differential equation of Eq. (\ref{covariant NPDE for U}). Notice that the spacetime support of $\rho_0(q)$ is  composed by $\mathcal{M}_{st}\equiv\mathcal{M}_s\otimes\mathcal{M}_t$. Inside $\mathcal{M}_{st}$, the quantum potential is thus flat given by 
\begin{equation}
U(q)=U_{s_0}+U_{t_0}=\frac{1}{2}\Big(\hbar^2k_0^2-\frac{\hbar^2\omega_0^2}{c^2}\Big)\equiv U_0. 
\label{quantum potential-energy-momentum}
\end{equation}
One therefore observes that at $T=0$, the quantum force is vanishing inside the spacetime support, $\partial^aU=0$. Below we shall proceed to consider only the case of vanishing $T$, $T=0$. 

Now, at $\tau=0$, let us choose the following initial wave function, $\{\rho_0(q),v_0^{\hspace{1mm}a}(q)\}$. Here $\rho_0(q)=\rho_{s_0}(r)\rho_{t_0}(t)$ and $v_0^{\hspace{1mm}a}(q)$ is a uniform velocity vector field having non-vanishing value only inside the spacetime support $\mathcal{M}_{st}$, that is $v_0^{\hspace{1mm}a}(q)=v_C^{\hspace{1mm}a}$, where $v_C^{\hspace{1mm}a}$ is a constant four velocity vector. Since at $\tau=0$ the quantum force is vanishing, then initially one has $dv^a/d\tau=0$. Hence, at infinitesimal lapse of proper time, $\tau=\Delta\tau$, the velocity field is kept uniform and constant. This in turn will shift the initial quantum probability density in spacetime by $\Delta q^{a}=v_C^{\hspace{1mm}a}\Delta\tau$, while keeping its profile unchanged: $\rho(q;\Delta \tau)=\rho_0(q^a-v_C^{\hspace{1mm}a}\Delta \tau)$. Accordingly, the spacetime support will also be shifted by the same amount. The same thing will happen for the next infinitesimal lapse of proper time and so on. Hence, at finite lapse of proper time, the velocity field is kept uniform and constant, $v^a(q;\tau)=v_0^{\hspace{1mm}a}(q^b-v_C^{\hspace{1mm}b}\tau)$. One therefore finally has:  $d_{\tau}\rho=\partial_{\tau}\rho+v^a\partial_a\rho=\partial_{\tau}\rho+\partial_a(\rho v^a)=0$, where in the second and third equalities we have used the uniformity of the velocity field inside the (continuously shifted) spacetime support and the continuity equation of Eq. (\ref{covariant Madelung fluid dynamics}). One can thus concludes that {\it the pair of fields  
\begin{equation}
\{\rho(q;\tau),v^a(q;\tau)\}=\{\rho_0(q^b-v_C^{\hspace{1mm}b}\tau),v_0^{\hspace{1mm}a}(q^b-v_C^{\hspace{1mm}b}\tau)\},
\end{equation}
comprises the stationary wave function of the relativistic Madelung fluid dynamics}. Here, $q$ belongs to the spacetime support at proper time $\tau$, denoted by  $\mathcal{M}_{st}^{\tau}$.

Next, since inside $\mathcal{M}_{st}^{\tau}$, the quantum potential is constant given by $U_0$, one has $U_{\mbox{\small tot}}=\int dq U\rho=U_0$. Let us proceed to choose a sufficiently large $\omega_0$ by picking sufficiently large $|U_{t_0}|$ so that $U_{\mbox{\small tot}}=U_0$ given in Eq. (\ref{quantum potential-energy-momentum}) is negative, $U_{\mbox{\small tot}}<0$. This allows us to define a new quantity $m$ as 
\begin{equation}
U_{\mbox{\small tot}}=U_0=-\frac{1}{2}m^2c^2, 
\label{emergent mass}
\end{equation}
so that inserting into Eq. (\ref{quantum potential-energy-momentum}) one obtains
\begin{equation}
\frac{\hbar^2\omega_0^2}{c^2}-\hbar^2k_0^2=m^2c^2.
\label{energy-momentum-mass relation}
\end{equation}
Recalling again the definition of quantum  potential of Eq. (\ref{covariant quantum potential}), Eq. (\ref{emergent mass}) can be put as 
\begin{equation}
\Box I(q;\tau)+\frac{m^2c^2}{\hbar^2}I(q;\tau)=0, \hspace{3mm}q\in\mathcal{M}_{st}^{\tau},
\label{Klein-Gordon equation}
\end{equation}
where $I\equiv\rho^{1/2}$. 

Now, let us give physical interpretation to the above formalism. First, Eq. (\ref{Klein-Gordon equation}) is but the Klein-Gordon equation with mass term $m$. One however should keep in mind that in contrast to the conventional Klein-Gordon equation, $\rho(q;\tau)=I^2(q;\tau)$ can be unambiguously regarded as probability density; and most importantly, the differential equation of Eq. (\ref{Klein-Gordon equation}) must be subjected to the boundary condition that $I(q;\tau)$ is vanishing at the surface boundary of the spacetime support, $\mathcal{M}_{st}^{\tau}$. One can also see that though $I(q)$ is not Lorentz invariant, $m$ is. 

Next, assuming that the de Broglie-Einstein relation is satisfied: $E=\hbar\omega_0$ and ${\bf p}^2=\hbar^2k_0^2$, where $E$ and ${\bf p}$ are classical energy and linear momentum respectively, then Eq. (\ref{energy-momentum-mass relation}) translates into 
\begin{equation}
\frac{E^2}{c^2}-{\bf p}^2=m^2c^2. 
\label{energy-momentum relation}
\end{equation}
The above is but the classical energy-momentum relation. Since one can choose $k_0$ and $\omega_0$ independently of ${\bf p}$ and $E$, then the above result opens the possibility of the violation of classical energy-momentum relation provided that the de Broglie-Einstein relation is not satisfied.  

Finally, at any physical time $t$, the stationary-moving spacetime wave function possesses {\it finite time uncertainty} given by: $\Delta_t\equiv 2t_0$. Using the boundary condition for temporal part of quantum probability density, one has 
\begin{equation}
\Delta_t=\frac{\pi}{\omega_0}=\frac{\pi\hbar}{c\sqrt{\hbar^2k_0^2+m^2c^2}}\equiv\frac{\pi\hbar}{\mathcal{E}}, 
\label{time uncertainty}
\end{equation}
where we have defined $\mathcal{E}$ as  $\mathcal{E}^2/c^2\equiv\hbar^2k_0^2+m^2c^2$. Notice that if the de Broglie-Einstein relation is satisfied then one has $\mathcal{E}=E$. Hence, time uncertainty decreases as one increases the mass term of the Klein-Gordon equation, $m$, and eventually vanishes for infinite $m$. In particular, if the de Broglie-Einstein relation holds,  then one has $\Delta_t=\pi\hbar/E$. 

{\it Conclusion}. --- To conclude, starting from relativistic mass-less Madelung fluid dynamics, we have developed a class of spatiotemporally-localized with finite-size spacetime support, uniformly moving thus stationary spacetime wave functions which maximizes Shannon entropy given its average quantum potential. The last aspect of the wave function allows us to regard it as {\it typical}. It turns out that the quantum amplitude of the wave function satisfies Klein-Gordon equation with emergent mass term $m$ proportional to the square root of the average quantum potential. We also rederived the classical energy-momentum relation provided that the de Broglie-Einstein relation is satisfied. It is then tempting to interpret the stationary-moving typical wave function as describing {\it a particle with a given mass $m$}. 

Further, we showed that at any physical time $t$, the stationary-moving spacetime wave function possesses {\it finite time uncertainty}. Namely, the temporal part of the quantum probability density extends while decays towards finite past and future. We showed that the physical time uncertainty decreases as $m$ increases and eventually vanishes for infinite $m$. In particular, if the de Broglie-Einstein relation holds, then the physical time uncertainty is proportional to the inverse of classical energy. 

\begin{acknowledgments}
This work is supported by FPR program at RIKEN. 
\end{acknowledgments}

\end{document}